# Semileptonic Decays of Pseudoscalar Mesons: A New Approach

Swee Ping Chia

*Physics Department, University of Malaya*
*50603 Kuala Lumpur, Malaysia*
*spchia@um.edu.my*

**Abstract.** Because quarks in hadronic states are subjected to strong QCD forces, it is not straight-forward to apply calculation obtained at the quark level to physical processes involving hadrons. In this paper we present a new approach to treat the semileptonic decays of pseudoscalar mesons of type $M_1 \to M_2 + l + \bar{\nu}$, where $M_1$ and $M_2$ are pseudoscalar mesons with quark contents of $q_1 \bar{q}$ and $q_2 \bar{q}$ respectively. At the quark level, the process $q_1 \to q_2 + l + \bar{\nu}$ is straightforward. However, when we fold the quark-level process to the meson decay process, the complication from QCD effects come in. We make the fundamental assumption that the vertex of type $Mq\bar{q}$ can be approximated by an effective constant $\gamma_5$ coupling. With this assumption, the hadronic process can then be related to the quark-level process. The model is applied to the semileptonic decays of $\pi$, $K$, $D$ and $B$ mesons.

## INTRODUCTION

The Standard Model (SM) has been successful in providing good agreement with experimental data. Many features of the SM have been well tested. However, although SM provides a general framework for quarks and leptons, it does not offer easy means for calculation when it comes to processes that involve strong QCD effects. Even for electroweak processes, although relatively simple at the quark level, the calculational details become involved and complicated when the processes are folded into hadronic states. This is because quarks are tightly bound inside hadrons. Because of the non-perturbative nature of such QCD forces, the description of the interactions of bound quarks in hadrons is still questionable.

In this paper, I shall focus on weak decay processes. Weak decay of quark represented by $q_1 \to q_2 \ell \nu$ is a simple process. How to relate the quark process to the corresponding hadron process $M_1 \to M_2 \ell \nu$ is, however, not straightforward. A simple model is proposed here to describe how meson is coupled to quarks. For simplicity, we shall consider only pseudoscalar mesons in here. Our model is based on the assumption that the meson-quark coupling can be adequately described by an effective constant $\gamma_5$ coupling. Such an approximation was successful in the calculation of the charge radii of pion and kaon [1], and in the calculation of the slope and magnitude of $f_\pi$ [2-5]. This approximation has also been employed in the calculation of $\bar{K}^0 \to$ vacuum amplitude in relation to $\Delta I = 1/2$ rule [6,7].

# SEMI-LEPTONIC DECAY OF MESON

The diagrammatic representation of semi-leptonic decay of pseudoscalar meson is as shown in Fig. 1. The amplitude for the diagram in Fig. 1 is given by

$$M = \frac{g^2}{2} g_1 g_2 V_{12} \left[ \int \frac{d^4k}{(2\pi)^4} \frac{N^\mu}{D} \right] \frac{-g_{\mu\nu} + q_\mu q_\nu / M_W^2}{q^2 - M_W^2} \bar{u}(p) \gamma^\nu L v(q-p) \quad (1)$$

where $g_1$ and $g_2$ are meson-quark couplings at $M_1$ and $M_2$ vertices, $V_{12}$ is the CKM mixing matrix element, and $k$, $q$, $p$ are the momenta of the quark $q_1$, the W-boson, and the charged lepton respectively. $N^\mu$ and $D$ are respectively:

$$N^\mu = Tr\left[\gamma_5 (\slashed{P} - \slashed{k} - m)\gamma_5 (\slashed{k} - \slashed{q} - m_2)\gamma^\mu L(\slashed{k} - m_1)\right] \quad (2)$$

$$D = [(P-k)^2 - m^2][(k-q)^2 - m_2^2][k^2 - m_1^2] \quad (3)$$

In the above expressions, $P$ is the momentum of the decaying meson $M_1$. The $\gamma_5$ couplings at the meson-quark vertices are explicitly displayed in the expression for $N^\mu$.

The integration over the internal momentum $k$ is logarithmically divergent. A cut-off momentum $\Lambda$ is introduced to tame the divergence, which yields, after some algebra

$$\int \frac{d^4k}{(2\pi)^4} \frac{N^\mu}{D} = 2\int_0^1 dx \int_0^{1-x} dy \frac{i}{32\pi^2} \left(\frac{3}{2} P^\mu\right) \ln \frac{\Lambda^2}{M^2} \quad (4)$$

For simplicity, we assume $\Lambda/M$ to be sufficiently large.

Putting Eq. (4) into Eq. (1), summing over the final states, and integrating over the phase space of the final states, the following expression is obtained for the decay rate:

$$\Gamma = \frac{9}{256\pi^3} G_F^2 \left[g_1 g_2 \ln(\Lambda^2/M^2)\right]^2 |V_{12}|^2 M_1^5 K(\varepsilon) \quad (5)$$

Here, $G_F$ is the Fermi coupling constant given by

$$G_F = \frac{g^2}{4\sqrt{2} M_W^2} = 1.1663787 \times 10^{-11} MeV^{-2} \quad (6)$$

$$K(\varepsilon) = -\varepsilon(1-\varepsilon)^2 \ln(1-\varepsilon) - \frac{1}{6}\varepsilon^2(6 - 9\varepsilon + 2\varepsilon^2) \quad (7)$$

$$\varepsilon = 1 - M_2^2 / M_1^2 \quad (8)$$

## Comparison with Experimental Values

It is observed from Eq. (5) that the decay rate depends on: (i) the overall coupling $G_F$, (ii) combined meson-quark coupling $g_1 g_2 \ln(\Lambda^2/M^2)$, (iii) CKM mixing matrix element $V_{12}$, and (iv) meson mass factor $M_1^5 K(\varepsilon)$. The CKM mixing matrix is given by [8]

$$|\mathbb{V}| = \begin{pmatrix} V_{ud} & V_{us} & V_{ub} \\ V_{cd} & V_{cs} & V_{cb} \\ V_{td} & V_{ts} & V_{tb} \end{pmatrix} = \begin{pmatrix} 0.97425 & 0.2252 & 0.00393 \\ 0.230 & 0.953075 & 0.0406 \\ 0.0084 & 0.0387 & 0.9992 \end{pmatrix} \quad (9)$$

When applying Eq. (5) to physical meson decay processes, a further factor $\kappa$ has to be considered. For the decay $\pi^+ \to \pi^0 e^+ \nu_e$, because $|\pi^+\rangle = |u\bar{d}\rangle$ and $|\pi^0\rangle = |(u\bar{u} - d\bar{d})/\sqrt{2}\rangle$, the decay can proceed via the decay of $u$ to $d$ or $\bar{d}$ to $\bar{u}$. A factor of $\kappa = 2$ is introduced. For the decay $K^+ \to \pi^0 e^+ \nu_e$, on the other hand, because $|K^+\rangle = |u\bar{s}\rangle$, the decay can only proceed via the decay of $\bar{s}$ to $\bar{u}$, and the additional factor is $\kappa = ½$. This factor of ½ is to be added for the decays $D^+ \to \pi^0 e^+ \nu_e$ and $B^+ \to \pi^0 \ell^+ \nu_\ell$ as well.

The decay rates can be deduced from the corresponding branching ratios [8] for each of the decay processes considered. Table 1 displays the decay rates for a number of decay modes. Table 2 shows the factor $\kappa$ and kinematic factor $K(\varepsilon)$ and the corresponding values of $g_1 g_2 \ln(\Lambda^2 / M^2)$ as deduced from Eq. (5).

**Conclusion**

What we have presented is a simple model to describe the hadronic electroweak processes in terms of processes at the quark level. The model is based on the assumption that the meson-quark coupling can be adequately described by an effective constant $\gamma_5$ coupling. In applying the model to semileptonic weak decays of pseudoscalar mesons, we obtain a consistent set of values for the parameter $g_1 g_2 \ln(\Lambda^2 / M^2)$. The decay of π-meson yields a value of about 0.4, whereas for the decay of K-meson, a value of about 0.5 is obtained. The decay of D-mesons yields a value of about 0.4. The decays of B-meson into π-meson gives a value of about 0.2, but a value of about 0.35 when it decays into D-meson. Overall, the values for the parameter $g_1 g_2 \ln(\Lambda^2 / M^2)$ are sufficiently close, indicating that the model is consistent, and the basic assumption is pointing to the right direction.

### ACKNOWLEDGMENTS

This research was carried out under the Research Grant UMRG049/09AFR from University of Malaya. Part of this research was carried while on research leave at Academia Sinica, Taiwan. The author wishes to thank Prof. Hai-Yang Cheng for valuable discussion.

**FIGURE 1.** Diagram representing $M_1 \to M_2 \ell \nu$ in this model, where $M_1$ and $M_2$ are $q_1 \bar{q}$ and $q \bar{q}_2$ respectively.

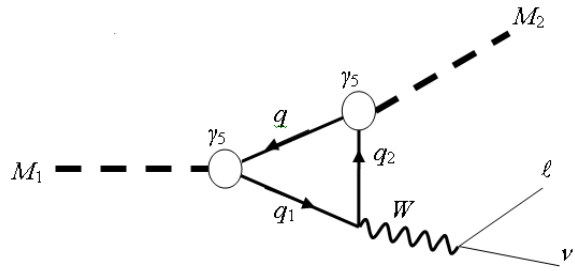

**Table 1.** Semi-leptonic decay rates of mesons

| Particle | Mean Life (s) | Decay Modes | Branching Ratio | Decay Rate (MeV) |
|---|---|---|---|---|
| $\pi^+$ | $2.6033(5) \times 10^{-8}$ | $\pi^0 e^+ \nu_e$ | $1.036(6) \times 10^{-8}$ | $2.619(15) \times 10^{-22}$ |
| $K^+$ | $1.2380(21) \times 10^{-8}$ | $\pi^0 e^+ \nu_e$ | $5.07(4)\,\%$ | $2.696(21) \times 10^{-15}$ |
| $K^0_L$ | $5.116(21) \times 10^{-8}$ | $\pi^- e^+ \nu_e$ | $40.55(11)\,\%$ | $5.217(14) \times 10^{-15}$ |
| $D^+$ | $1.040(7) \times 10^{-12}$ | $\pi^0 e^+ \nu_e$ | $4.05(18) \times 10^{-3}$ | $2.56(11) \times 10^{-12}$ |
|  |  | $\overline{K}^0 e^+ \nu_e$ | $8.83(22)\,\%$ | $5.588(14) \times 10^{-11}$ |
| $D^0$ | $0.4101(15) \times 10^{-12}$ | $\pi^- e^+ \nu_e$ | $2.89(8) \times 10^{-3}$ | $4.64(13) \times 10^{-12}$ |
|  |  | $K^- e^+ \nu_e$ | $3.55(4)\,\%$ | $5.70(6) \times 10^{-11}$ |
| $B^+$ | $1.641(8) \times 10^{-12}$ | $\pi^0 \ell^+ \nu_\ell$ | $7.78(28) \times 10^{-5}$ | $3.12(11) \times 10^{-14}$ |
|  |  | $\overline{D}^0 \ell^+ \nu_\ell$ | $2.26(11)\,\%$ | $8.9(4) \times 10^{-12}$ |
| $B^0$ | $1.519(7) \times 10^{-12}$ | $\pi^- \ell^+ \nu_\ell$ | $1.44(5) \times 10^{-4}$ | $6.24(22) \times 10^{-14}$ |
|  |  | $D^- \ell^+ \nu_\ell$ | $2.18(12)\,\%$ | $9.5(5) \times 10^{-12}$ |

*Data are taken from J. Beringer *et al*. (Particle Data Group) [Ref. 8].

**TABLE 2.** Values for $g_1 g_2 \ln(\Lambda^2/M^2)$ obtained from semi-leptonic decay rates.

| Decay Mode | Decay Rate (MeV) | $\kappa$ | $K(\varepsilon)$ | $g_1 g_2 \ln(\Lambda^2/M^2)$ |
|---|---|---|---|---|
| $\pi^+ \to \pi^0 e^+ \nu_e$ | $2.619(15) \times 10^{-22}$ | 2 | $0.97320 \times 10^{-7}$ | 0.4166 |
| $K^+ \to \pi^0 e^+ \nu_e$ | $2.696(21) \times 10^{-15}$ | ½ | $0.10117$ | 0.4820 |
| $K^0_L \to \pi^- e^+ \nu_e$ | $5.217(14) \times 10^{-15}$ | 1 | $0.98573 \times 10^{-1}$ | 0.4709 |
| $D^+ \to \pi^0 e^+ \nu_e$ | $2.56(11) \times 10^{-12}$ | ½ | $0.16078$ | 0.4133 |
| $D^+ \to \overline{K}^0 e^+ \nu_e$ | $5.588(14) \times 10^{-11}$ | 1 | $0.10382$ | 0.4101 |
| $D^0 \to \pi^- e^+ \nu_e$ | $4.64(13) \times 10^{-12}$ | 1 | $0.16034$ | 0.3965 |
| $D^0 \to K^- e^+ \nu_e$ | $5.70(6) \times 10^{-11}$ | 1 | $0.10435$ | 0.4158 |
| $B^+ \to \pi^0 \ell^+ \nu_\ell$ | $3.12(11) \times 10^{-14}$ | ½ | $0.16591$ | 0.1962 |
| $B^+ \to \overline{D}^0 \ell^+ \nu_\ell$ | $8.9(4) \times 10^{-12}$ | 1 | $0.72401 \times 10^{-1}$ | 0.3434 |
| $B^0 \to \pi^- \ell^+ \nu_\ell$ | $6.24(22) \times 10^{-14}$ | 1 | $0.16586$ | 0.1962 |
| $B^0 \to D^- \ell^+ \nu_\ell$ | $9.5(5) \times 10^{-12}$ | 1 | $0.72097 \times 10^{-1}$ | 0.3554 |